\pdfoutput=1

\documentclass[twocolumn]{aastex63}
\usepackage[normalem]{ulem}
\usepackage{bm}
\usepackage[utf8]{inputenc}
\usepackage{amssymb}
\usepackage{amsthm}
\usepackage{amsmath}
\usepackage{mathrsfs}

\begin{document} 
\title{Search for Black Hole Merger Families}
\correspondingauthor{Do\u{g}a Veske}
\email{dv2397@columbia.edu}

\author[0000-0003-4225-0895]{Do\u{g}a Veske}
\affiliation{Department of Physics, Columbia University in the City of New York, New York, NY 10027, USA}
\author[0000-0002-9545-7286]{Andrew G. Sullivan}
\affiliation{Department of Physics, Columbia University in the City of New York, New York, NY 10027, USA}
\author[0000-0003-1306-5260]{Zsuzsa M\'arka}
\affiliation{Columbia Astrophysics Laboratory, Columbia University in the City of New York, New York, NY 10027, USA}
\author[0000-0001-5607-3637]{Imre Bartos}
\affiliation{Department of Physics, University of Florida, PO Box 118440, Gainesville, FL 32611-8440, USA}
\author{K. Rainer Corley}
\affiliation{Department of Physics, Columbia University in the City of New York, New York, NY 10027, USA}
\affiliation{Columbia Astrophysics Laboratory, Columbia University in the City of New York, New York, NY 10027, USA}
\author[0000-0003-0607-8741]{Johan Samsing}
\affiliation{Niels Bohr International Academy, The Niels Bohr Institute, Blegdamsvej 17, DK-2100, Copenhagen, Denmark}
\author[0000-0002-7387-6754]{Riccardo Buscicchio}
\affiliation{Institute for Gravitational Wave Astronomy 
\& School of Physics and Astronomy, University of 
Birmingham, Birmingham, B15 2TT, UK}
\author[0000-0002-3957-1324]{Szabolcs M\'arka}
\affiliation{Department of Physics, Columbia University in the City of New York, New York, NY 10027, USA}

\begin{abstract}
The origin, environment, and evolution of stellar-mass black hole binaries are still a mystery. One of the proposed binary formation mechanisms is manifest in dynamical interactions between multiple black holes. A resulting framework of these dynamical interactions is the so-called hierarchical triple merger scenario, which happens when three black holes become gravitationally bound, causing two successive black hole mergers to occur. In such successive mergers, the black holes involved are directly related to each other, and hence this channel can be directly tested from the properties of the detected binary black hole mergers. Here we present a search for hierarchical triple mergers among events within the GWTC-1 and GWTC-2 catalogs of LIGO/Virgo, the eccentric localization of GW190521 and those found by the IAS-Princeton group. The search includes improved statistical quantification that also accounts for black hole spins.  We perform our analysis for different upper bounds on the mass distribution of first generation BHs. Our results demonstrate the importance of the mass distributions' properties for constraining the hierarchical merger scenario. We present the individually significant merger pairs. The search yields interesting candidate families and hints of its future impact.
\end{abstract}

\keywords{Astrophysical black holes (98) -- Gravitational wave astronomy (675) -- Gravitational wave sources (677)}

% -------------------------------------------
\section{Introduction}
\label{sec:Introduction}
With the gravitational-wave (GW) detectors LIGO \citep{Aasi_2015} and Virgo \citep{Acernese_2014} reaching necessary sensitivities for recognizing binary black hole (BBH) merger signals  \citep{2016PhRvL.116f1102A}, an increasing number of detected BBH mergers are being collected. The GW strain data has been analyzed by LIGO Scientific and Virgo Collaborations and GW catalogues have been released \citep{Abbott_2019,2020arXiv201014527A}.
Since the data was made public  \citep{theligoscientificcollaboration2019open}, individual groups have found additional BBH mergers as well \citep{venumadhav2019new,zackay2019detecting,Nitz_2019,Zackay_2019}. 
As the number of detections increases, a wider array of unique mergers are detected which provide tests of general relativity and its alternatives at different conditions \citep{2019PhRvD.100j4036A,theligoscientificcollaboration2020tests}

Another piece of information that can be acquired from gravitational wave bursts relates to the interactions of black holes (BH) with each other as there are various proposed formation channels for inspiraling BBH systems~\citep{2016ApJ...832L...2R, 2017ApJ...846...82Z,PhysRevLett.123.181101, 2014ApJ...784...71S, 2017ApJ...840L..14S, 2018ApJ...853..140S, 2018PhRvD..97j3014S, 2018ApJ...855..124S, 2018MNRAS.tmp.2223S, 2019ApJ...871...91Z, 2018PhRvD..98l3005R, 2019arXiv190711231S, 2019PhRvD.100d3010S,vignagomez2020massive,Fragione_2019}. 
Available formation scenarios include field binaries~\citep{2012ApJ...759...52D, 2013ApJ...779...72D, 2015ApJ...806..263D, 2016ApJ...819..108B,2016Natur.534..512B, 2017ApJ...836...39S, 2017ApJ...845..173M, 2018ApJ...863....7R, 2018ApJ...862L...3S}, 
chemically homogeneous binary evolution~\citep{10.1093/mnras/stw1219,10.1093/mnras/stw379,refId0}, 
dynamical mergers in dense stellar clusters~\citep{2000ApJ...528L..17P,2010MNRAS.402..371B, 2013MNRAS.435.1358T, 2014MNRAS.440.2714B,2015PhRvL.115e1101R, 2016PhRvD..93h4029R, 2016ApJ...824L...8R,2017MNRAS.464L..36A, 2017MNRAS.469.4665P, 2018PhRvD..97j3014S, 2018MNRAS.481.5445S}, 
in active galactic nuclei (AGN) discs~\citep{2017ApJ...835..165B,2017NatCo...8..831B,2018ApJ...866...66M,Gayathri_2020,Graham_2020,2019ApJ...876..122Y,2019MNRAS.488.4459C,2019ApJ...884L..50M,2017MNRAS.464..946S,2012MNRAS.425..460M,2014MNRAS.441..900M,2016ApJ...819L..17B,2020arXiv201009765S,PhysRevLett.123.181101,2020ApJ...900L..13A,2020ApJ...896..138Y}
and in galactic nuclei (GN)~\citep{2009MNRAS.395.2127O, 2015MNRAS.448..754H,2016ApJ...828...77V, 2016ApJ...831..187A, 2016MNRAS.460.3494S, 2017arXiv170609896H, 2018ApJ...865....2H,2019MNRAS.488...47F}, 
very massive stellar mergers~\citep{Loeb:2016, Woosley:2016, Janiuk+2017, DOrazioLoeb:2017} 
and single-single GW captures of primordial black holes~\citep{2016PhRvL.116t1301B, 2016PhRvD..94h4013C,2016PhRvL.117f1101S, 2016PhRvD..94h3504C}.

\begin{figure} [hbt!]
    \centering
    \includegraphics[width=\columnwidth]{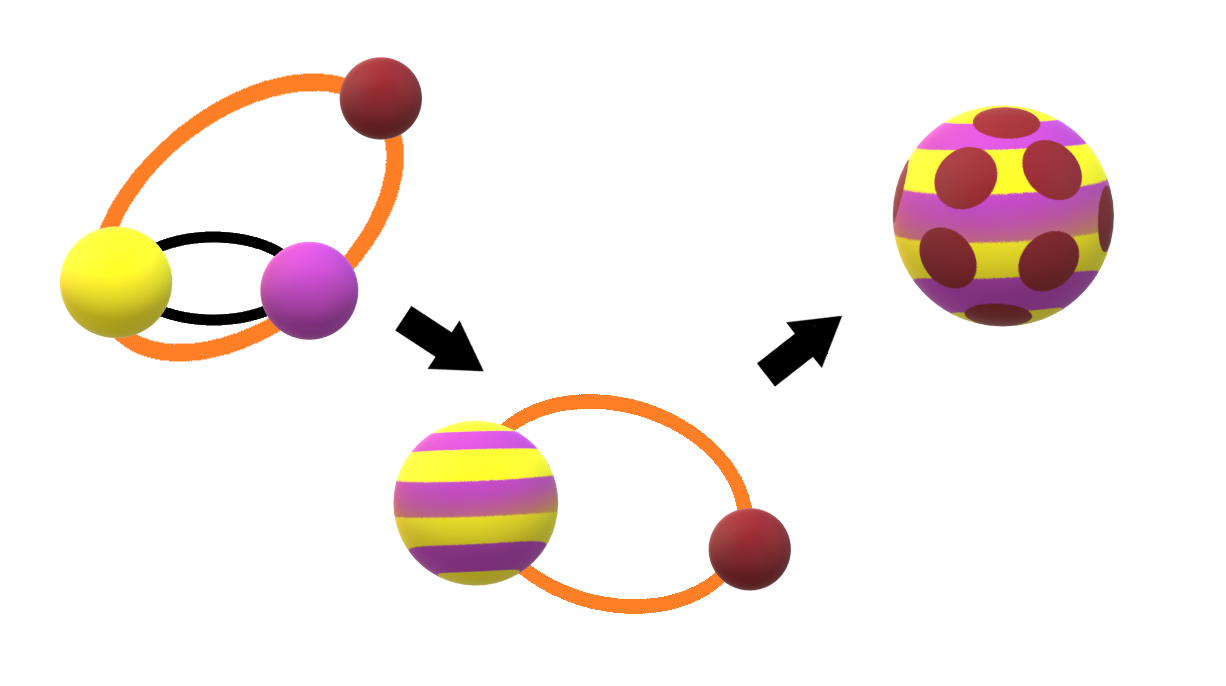}
    \caption{Depiction of a triple hierarchical merger, for a bound 3 BH system. We show the relationship between the generation of BHs with their colors. Arrows indicate the chronological order.}
    \label{fig:triple}
\end{figure}

One specific BBH merger scenario of interest is the so-called hierarchical triple merger, which is proposed to occur in the dynamical BBH formation channels \citep{10.1093/mnras/sty197, 10.1093/mnras/sty2249}. 
Hierarchical triple mergers can occur through the interaction of three BHs $(BH_1,BH_2,BH_3)$ which form a gravitationally bound three-body system. 
This three-body interaction facilitates the inspiral of the BHs in the system \citep{Campanelli:2007ea,Lousto:2007rj}. First two BHs merge emitting a GW signal, leaving behind a BBH composed of the first merger remnant $(BH_1,BH_2 \rightarrow BH_{12})$. 
It eventually merges with the remaining single BH $(BH_3)$ to produce a second GW signal $(BH_{12},BH_3 \rightarrow BH_{123})$ \citep{10.1093/mnras/sty2249}. The process is depicted in Fig. \ref{fig:triple}

Under certain orbital configurations, it has been shown that both mergers can be observed within timescales of about few years \citep{10.1093/mnras/sty2249}. 
In previous work \citep{2020MNRAS.tmpL.133V}, this scenario has been observationally constrained by using the mergers in the first GW transient catalog (GWTC-1) \citep{Abbott_2019} and the mergers found by the IAS-Princeton group \citep{venumadhav2019new,zackay2019detecting,Zackay_2019} from LIGO/Virgo's first and second observing runs (O1 and O2).

In this paper, we extend our previous search \citep{2020MNRAS.tmpL.133V} by including the BBH mergers (except the single detector detection GW190424) from the second GW transient catalog (GWTC-2)~\citep{2020arXiv201014527A} published by LIGO Scientific Collaboration and Virgo Collaboration which includes the mergers from the first half of LIGO and Virgo detectors' third observing run (O3a) and the eccentric localization of GW190521 \citep{gayathri2020gw190521}. In addition, we improve our test statistic by using the BH spins estimated for each merger. 

This Letter is organized as follows; in Sec. \ref{sec:search} we describe the details of our search. In Sec. \ref{sec:res} we present our results and provide discussion. In Sec. \ref{sec:conc} we conclude.

\section{Search}
\label{sec:search}
Our search analyzes pairs of BBH mergers, using a number of parameters which behave differently under the two hypotheses under consideration: the two mergers are unrelated and the two mergers are related through the hierarchical triple merger scenario. For hierarchical triple mergers, the mass and spin of the remnant BH from the first merger must correspond to the mass and spin of one of the BHs in the second merger. 
In addition, the sky localizations of the first and second mergers must be overlapping. In the case of unrelated mergers, the location of two mergers and the masses and spins of their component BHs are independent of each other. 
Our search is based on a frequentist $p$-value assignment through the use of a test statistic (TS).
As Neyman-Pearson's lemma suggests \citep{doi:10.1098/rsta.1933.0009}, we choose our TS to be the likelihood ratio of the \emph{signal} hypothesis \(H_s\) -- a hierarchical triple merger, and the \emph{null} hypothesis \(H_0\) - two unrelated mergers.

\subsection{Merger Properties}
\label{sec:prop}
Let merger \#1 be the merger whose remnant then participates in the second merger, which will be denoted as merger \#2. We use four properties of each BBH merger for calculating the likelihood ratios, which we list below. 
\begin{itemize}
    \item {\it Correct time order:} For hierarchical triples, merger \#1 must occur before merger \#2.
    \item {\it Mass estimates:}  The mass of the remnant of merger \#1 should be consistent with the mass of one of the BHs in merger \#2. For unrelated mergers the masses are assumed to be independently drawn from the initial BH mass distribution. Both theoretical and empirical estimates of the initial BH mass distributions are uncertain. Therefore we consider several different mass distributions as explained in Sec. \ref{sec:mass}.
    \item {\it Effective spin parameter $\chi_\text{eff}$:}  For assessing consistency between the remnant of merger \#1 and the BHs in merger \#2, beyond mass we also consider the BHs spin. Remnants from BBH mergers are expected to be highly spinning, typically around the dimensionless spin magnitude $|\vec{a}|=|c\vec{S}/Gm^2|\sim 0.7$ \citep{Berti_2008}, where $c$ is the speed of light, $\vec{S}$ is the spin angular momentum of the BH, $G$ is the universal gravitational constant and $m$ is the mass of the BH.
    For spin comparison we use the effective spin parameter $\chi_\text{eff}$ which is the best constrained by gravitational waves measurements. We use $\chi_\text{eff}$ instead of the actual spins due to the fact that parameter estimations of $\chi_\text{eff}$ can be approximated as normal distributions, where as spin distributions cannot be approximated with a simple parameterization. We elaborate on it more in Sec. \ref{bg}.  $\chi_\text{eff}$ is defined as 
    \begin{equation}
        \chi_\text{eff} = \frac{\Vec{a_1}m_1+\Vec{a_2}m_2}{m_1+m_2}\cdot \hat{L}
    \end{equation}
    where $m_{1,2}$ are the masses of the initial BHs in the merger, $\Vec{a_{1,2}}$ are their dimensionless spin vectors and $\hat{L}$ is the unit vector of the binary's orbital angular momentum. 
    We assume an initial BH dimensionless spin magnitude to be uniform between [0,1]. We further assume that in BBH mergers spin orientations are random following a uniform isotropic distribution, which is expected for dynamical interactions \citep{Vitale_2017}, but not from isolated binaries (e.g. \cite{Bogdanovi__2007}).
    \item {\it Localization:} The localizations of mergers \#1 and \#2 must be consistent with each other. We assume them to happen at the same point in space and neglect possible travel between the mergers. We assume that the merger rate is uniform in comoving volume.
    \item {\it Eccentricity:} In triple hierarchical merger scenario, especially for merger \#1, eccentric mergers are expected \citep{10.1093/mnras/sty2249}. However, we do not use the eccentricity in our test statistic due to the following reasoning. The most sensitive BBH merger searches are template based matched filter searches which use circular orbits \citep{2020arXiv201014527A}, except the very recent work targeted on GW190521 \citep{gayathri2020gw190521}. Unmodeled burst searches can catch the eccentric mergers but with a lower reach; however, no such merger has been detected by unmodeled searches with high significance \citep{LIGOeccentric}. Due to the lack of sufficient eccentric waveforms and their computationally expensive analysis, the precise estimation of the eccentricity of the detected mergers cannot be done too. Nevertheless, there are eccentricity estimations for the detected mergers assuming low eccentricity, and events in GWTC-1 were found to have low eccentricity ($<0.1$) \citep{Romero_Shaw_2019}. On the other hand, the event GW190521 is claimed to have high eccentricity by individual parties \citep{gayathri2020gw190521,Romero_Shaw_2020}. We add the localization of the best matching eccentric template of GW190521 \citep{gayathri2020gw190521} to our analysis, as the only available eccentric localization.
\end{itemize}

\subsection{BH mass distributions}
\label{sec:mass}
An important part of the analysis is the mass distribution for first generation BHs. Although there is no constraint on the mass of BHs from general relativity, considering the formations channels from stellar evolution, a restricted region (known as the upper mass gap) appears between $\sim 50-135 M_{\odot}$ for first generation BHs due to phenomena known as pair instability supernova (PISN) and pulsational pair instability supernova (PPISN) \citep{Woosley_2017}.
Both are related to the process of creation of electron-positron pairs from the high pressure in the star which results in the reduction of the photon pressure, consequently an initial collapse and an explosion.
Under certain conditions, the star can survive with lighter mass, outside the forbidden region, after losing its mass partially by expelling its outer layers during series of explosions. This is called PPISN \citep{Yoshida_2016}. 
Conversely, after PISN, no compact remnant is left behind. Therefore finding a BH with a mass in the forbidden region may indicate having a higher generation BH which could not have originated directly from the stellar evolution \citep{2020arXiv200609744S}. Besides the higher generation BH scenario, such cases may also be considered under, but not limited to, primordial BHs or BHs involving particles beyond the standard model such as dark matter \citep{2016PhRvL.116t1301B,Clesse_2017}.

However, theoretical and numerical studies show that the bounds are not precise and also can vary with the initial composition of the star~\citep{Woosley_2017}. Besides the limits on the upper mass gap, the shape of the first generation BH mass distribution is not observationally well constrained and there are few appropriate mass distributions. However, the most recent estimates favor a power law+peak model the most~\citep{2020arXiv201014533T}. The peak is thought to carry the BHs which survived a PPISN.

Due to all of these uncertainties, we perform our search using different mass distributions separately. We adopt a single parametrization which is the power law+peak model in~\citep{2020arXiv201014533T}; but with different upper bounds~$=\{50,60,70,100\}M_{\odot}$. Except the upper bound, we use the fitted parameters for the power law+peak model in~\cite{2020arXiv201014533T}. We obtain the second generation BHs' mass distribution by adding the primary and secondary masses of the \#1 mergers as random variables, using their joint distribution. Finally, since we are interested in detected mergers, we modify the mass distributions by multiplying by $m_{secondary}^{1.5}$ which accounts for the detection bias. This approximate mass factor favoring heavier masses for detection comes analytically when imposing a threshold on signal-to-noise ratio (SNR) for detections for the uniformly distributed sources in space. Although the detection threshold is the false-alarm rate not the SNR in reality because of the non-gaussian Poisson-like noise called glitches, it is well correlated with SNR especially after glitch involving data parts are removed~\citep{ligonoise}.

\subsection{The background distribution}
\label{bg}
Our significance test is based on a frequentist $p$-value assignment via comparison with a background distribution. The background distribution involves simulated first generation BH mergers randomly matched with each other, which mimic the set of unrelated merger pairs. The background distribution generation algorithm is based on performing BBH merger simulations and localizing them with the BAYESTAR software package \citep{PhysRevD.93.024013,Singer_2016}. BAYESTAR is a fast localization software for GW detection from compact binary coalescences which takes $\mathcal{O}(1)$ minute to localize a merger by a central processing unit (CPU) core with a frequency of $\sim 1$ GHz. It was used for the fast-response skymaps for open public alerts during O3 which were used by multi-messenger searches.
It is  slightly less accurate than the localization from the computationally expensive full parameter estimation \citep{PhysRevD.93.024013}. However the full parameter estimation may take several days to finish in the same configuration per merger and hence is not appropriate for a background distribution generation. We construct distributions at 2 sensitivities, O1/O2 and O3 sensitivity, and for relevant detector combinations in the runs, considering 2 LIGO \citep{Aasi_2015} (Hanford and Livingston) and the Virgo \citep{Acernese_2014} detectors.

In order to construct the background distributions for our test statistic, we need the same inputs as real detections which have parameter estimations with variances. We model joint distributions of component masses of the second merger as bivariate normal distributions truncated such that the larger mass is always greater than the smaller one. The mean location of the distributions are chosen as the injected actual masses. The elements of the covariance of the distribution are fitted as a linear function of the detection signal-to-noise ratio (SNR). The parameters of the fit is extracted from the distributions of the real detections of GWTC-2. Similarly, we model the distributions of final mass and $\chi_\text{eff}$ of the first merger as normal distributions, whose variances are found from linear fits of SNR. For $\chi_\text{eff}$ we limit the distributions between [-1,1]. In accordance with our assumption for BHs involved in unrelated mergers (spin magnitude uniform in [0,1] and random orientations) we assign a $\chi_\text{eff}$ value to each merger. This $\chi_\text{eff}$ value and the 5\% reduced total mass of the merger constitute the mean location of the distributions.

Before moving on the results of our search, in order to estimate the possible capability of our search, we performed simulations for the triple hierarchical merger scenario and found that if two mergers of such an interaction are detected by 2 LIGO and the Virgo detectors (3 detector detection) at O3 sensitivity, they can be identified by our search as a triple hierarchical merger at $\sim90\%$ efficiency at $3\sigma$ confidence level ($p\lesssim$ 1/740).

\section{Results and Discussion}
\label{sec:res}
Here we present the results of our search. Table \ref{table1} shows the most significant event pairs and their individual $p$-values for different upper bounds of the first generation BHs' mass distribution. We also provide $p$-values for partial inputs: time, mass and spin only; and time and volume only. The $p$-values in Table \ref{table1} are for individual events and do not consider the multiple hypothesis testing correction. We discuss the multiple hypothesis testing correction at the end of this section.

\begin{table*}[]
\begin{center}

\begin{tabular}{|m{3.5cm}|m{3.5cm}|m{9cm}|}
\hline
\begin{tabular}[c]{@{}l@{}}Upper bound on 1G BH \\ mass  distribution [$M_{\odot}]$\end{tabular} & Most significant pairs   & \begin{center}\begin{tabular}[c]{@{}l@{}} Individual $p$-value: \\ Overall -- Time, mass and spin only -- Time and volume only \end{tabular}\end{center} \\ \hline
50                                                                          &    \begin{tabular}[c]{@{}l@{}}All pairs of GW190521 \\ GW190521-GW190514 \\ {GW190521-GW170823} \\GW190521-GW170729 \end{tabular}                  &    \begin{tabular}[c]{@{}l@{}} 0 (null hypothesis is  rejected with certainty) \\ 0 -- 0 -- {0.14 (0.14 for eccentric GW190521)}  \\ {0 -- 0 -- 0.45 (0.24 for eccentric GW190521)}\\0 -- 0 -- {0.32 (0.31 for eccentric GW190521)}\end{tabular}    \\ \hline
60                                                                          &   \begin{tabular}[c]{@{}l@{}} GW190519-GW170818 \\ GW190915-GW190708    \\  GW190910-GW190512    \end{tabular}         & \begin{tabular}[c]{@{}l@{}}  $1.1\times10^{-3}$ -- $4.9\times10^{-2}$ -- $3.9\times10^{-2}$\\$9.0\times10^{-3}$ -- $0.20$ -- $3.2\times10^{-2}$\\     
$1.3\times10^{-2}$ -- $0.25$ -- $1.8\times10^{-2}$\end{tabular} \\ 

\hline
70                                                                          &      \begin{tabular}[c]{@{}l@{}} GW190519-GW170818 \\ GW190910-GW190512 \\ GW190915-GW190708 
\end{tabular}                 &    \begin{tabular}[c]{@{}l@{}} $1.7\times10^{-3}$ -- $7.7\times10^{-2}$ -- $3.9\times10^{-2}$\\$8.1\times10^{-3}$ -- $0.26$ -- $1.8\times10^{-2}$\\     
$8.4\times10^{-3}$ -- $0.19$ -- $3.2\times10^{-2}$
\end{tabular} \\ \hline
100                                                                         &     \begin{tabular}[c]{@{}l@{}}   GW190519-GW170818 \\ GW190910-GW190512 \\ GW190915-GW190708 
\end{tabular}              &  \begin{tabular}[c]{@{}l@{}}  $1.5\times10^{-3}$ -- $5.6\times10^{-2}$ -- $3.9\times10^{-2}$\\$7.2\times10^{-3}$ -- $0.21$ -- $1.8\times10^{-2}$\\     
$7.5\times10^{-3}$ -- $0.18$ -- $3.2\times10^{-2}$
\end{tabular}  \\ \hline
\end{tabular}
\end{center}
\caption{The most significant pairs and their significances for the triple hierarchical merger scenario for different upper bounds for the mass distribution of first generation BHs assuming the power law+peak model. The given $p$-values do not include the multiple hypothesis correction factor.}
\label{table1}
\end{table*}

We find that for upper bound$=50 M_{\odot}$, GW190521 produces an infinite TS. This is caused by the fact that the mass estimation sample for the heavier component of GW190521 only involves masses greater than $50 M_{\odot}$. For the eccentric detection of GW190521, the parameters of the component masses for the best matching template are also well above $50 M_{\odot}$ at $102 M_{\odot}$ \citep{gayathri2020gw190521}. Therefore our null hypothesis, in this case, is rejected with certainty. One of the plausible explanations is it being a result of a previous merger. We provide the three significant pairs involving GW190521 with the partial search which uses only the time ordering and volume localizations: GW190521-GW190514, GW190521-GW170823 and GW190521-GW170729. The eccentric localization, which has a closer distance reconstruction, produces a substantially higher significance for GW190521-GW170823 pair. However, the parameters of the component masses of GW190521 for the best matching eccentric template is beyond the reach of the final mass of GW170823. Therefore without a new and proper parameter estimation for the eccentric analysis, GW170823 does not seem to be a plausible predecessor of GW190521 at the moment. Regarding  GW190521-GW190514 and GW190521-GW170729 about the triple hierarchical merger scenario, there are two points worth mentioning which make the pairs more notable:
\begin{itemize}
    \item The final masses of both GW190514 and GW170729 partially agree with the non-eccentric estimation and the best matching eccentric template's value of the heavier mass of GW190521.
    \item GW190514's 90\% sky localization is in agreement with the location of the AGN from which a candidate optical counterpart was detected after GW190521 by the Zwicky Transient Facility \citep{Graham_2020}. The distance reconstruction of the mergers are nearly identical along the AGN's direction.

\end{itemize}

\begin{figure}
    \centering
    \includegraphics[width=\columnwidth]{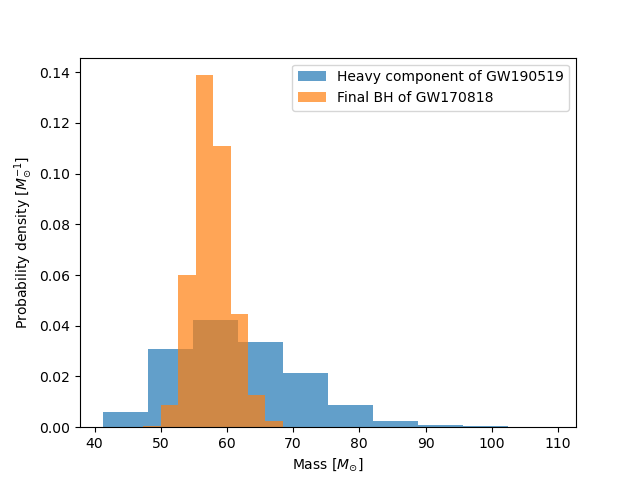}
    \caption{Probability densities for the mass estimations of the heavy component of GW190519 (blue) and the total mass of GW170818 reduced by 5\% for GW radiation (orange)}
    \label{fig:mass}
\end{figure}

\begin{figure}
    \centering
    \includegraphics[width=\columnwidth]{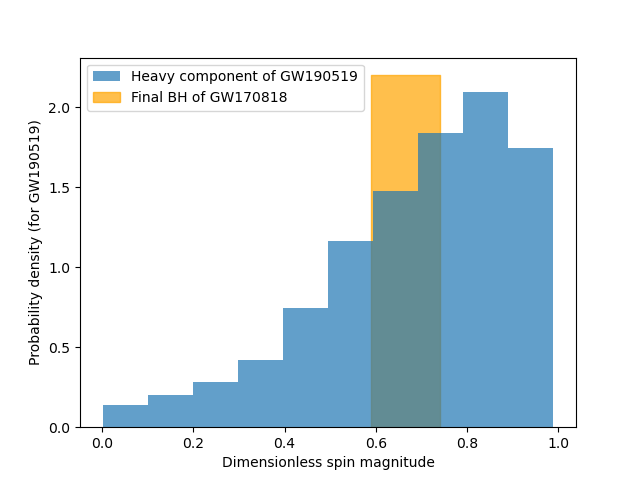}
    \caption{Probability density distribution of the dimensionless spin magnitude for the heavy component of GW190519 from its parameter estimation (blue) and range of it for the final BH of GW170818 with its uncertainties from GWTC-1\citep{Abbott_2019} (orange)}
    \label{fig:spin}
\end{figure}

\begin{figure*}
    \centering
    \includegraphics[width=\textwidth]{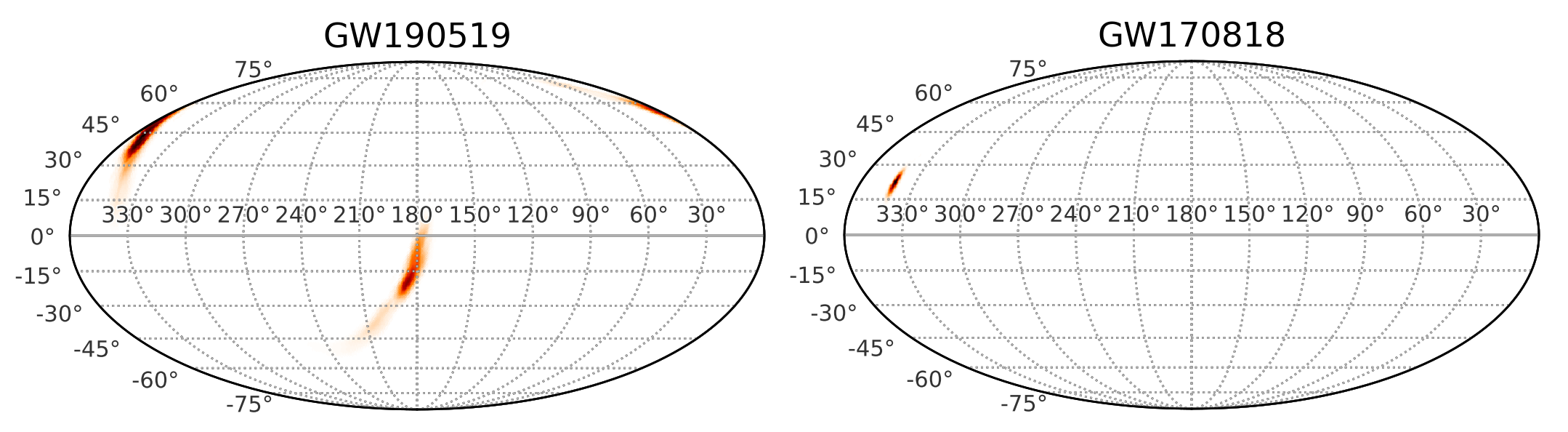}
    \caption{Sky localizations of GW190519 and GW170818 in equatorial coordinates. Darker color represents higher probability density.}
    \label{fig:skymap}
\end{figure*}

For 60, 70 and 100 $M_{\odot}$ for the mass upper bound, GW190519-GW170818 appear as the most significant pair with $\sim 3\sigma$ individual significance. When we investigate the pair's properties, we see that there is not a dominating single input parameter that makes the pair significant; but all the properties contribute to the significance. This can be seen from the $p$-values of the partial searches in Table \ref{table1}. We note that the distance reconstructions of two events seem to be peaked around different distances (2.85 and 1.06 Gpc); but with non-zero overlap. With a good 2D sky overlap, the overall volume overlap becomes non-negligible. Their joint distance reconstruction is peaked around 1.13 Gpc. The significance of the mass term comes from the overlap of the heavier mass of GW190519 ($m_1=64.5^{+11.3}_{-13.2}$) \citep{2020arXiv201014527A} with the final mass of GW170818 ($m_f=59.4^{+4.9}_{-3.8}$) \citep{Abbott_2019}, as seen in Fig. \ref{fig:mass}. In Fig. \ref{fig:spin} we show the parameter estimation of the spin of the heavy component of GW190519 and the final spin of GW170818 with the uncertainties provided in GWTC-1\citep{Abbott_2019} ($a_f=0.67^{+0.07}_{-0.08}$). We see that the GW190519's heavy component is expected to be a spinning BH whose dimentionless spin magnitude agrees with the spin of the final BH of GW170818. In our search due to limitations mentioned in Sec. \ref{sec:prop} we do not use individual spins; but the $\chi_\text{eff}$ parameter of the merger \#2. This choice does not directly favor spinning BHs. However, in this case, we do have a spinning BH which is expected from hierarchical merger scenarios and in particular the triple hierarchical merger scenario. We show the sky localizations of two mergers in Fig. \ref{fig:skymap}.

We also list GW190915-GW190708 and GW190910-GW190512 as the next two most significant pairs in Table \ref{table1} with significances $p \lesssim 1\%$. The pairs including GW190521 have relatively lower significances for higher mass upper bounds due to the relatively heavy secondary mass of GW190521 which does not fit well to the expected secondary masses for the triple hierarchical merger scenario. However, if GW190521 is considered a hierarchical merger, its heavier secondary mass can be explained by higher order hierarchical mergers (i.e. 2G+2G mergers \citep{Fragione_2019,Fragione_2020}).

In our test statistic we did not use the eccentricity of the mergers due to lack of analyses on the eccentricity of the mergers. However in the hierarchical merger scenario, the first merger especially is expected to have non-zero eccentricity \citep{10.1093/mnras/sty2249}. \cite{Romero_Shaw_2019} finds events in GWTC-1 have low eccentricity. This observation decreases the plausiblities of GW170818, GW170729 and GW170823 being the first merger of a triple hierarchical merger chain. For the events in GWTC-2, only GW190521 is analyzed for eccentricity and it is claimed to be consistent with $>0.1$ eccentricity \citep{gayathri2020gw190521,Romero_Shaw_2020}. This is specifically interesting for the GW190521-GW190514 pair, since two mergers have only one week in between. In such a short time period, the second merger would not have enough time to circularize and consequently is also expected to be eccentric. Therefore eccentric characteristics of GW190521 supports the possibility of GW190521-GW190514 pair to be a triple hierarchical merger chain. Further analysis of GW190514 can illuminate this claim more.

Finally, although we have individually significant events, it should be noted that for each upper bound on the mass distribution, we have analyzed 1431 pairs. Therefore a multiple hypothesis testing factor should be included for the analysis to bound the family-wise error rate, i.e. it is expected to have an pair with 1\% $p$-value if 100 unrelated pairs are analyzed by the definition of $p$-value. This does not change the interpretation for the upper bound=$50M_{\odot}$ case. For all the other limits, the search can only show the interesting merger pairs; but cannot provide a statistically significant pair with a low overall false alarm rate.

\section{Conclusion}
\label{sec:conc}
We presented a search for triple hierarchical mergers. We analyzed the events published in GWTC-1 \citep{Abbott_2019}, GWTC-2 \citep{2020arXiv201014527A}  (except the single detector detection GW190424), by the IAS-Princeton group \citep{venumadhav2019new,zackay2019detecting,Zackay_2019} and the eccentric localization of GW190521 \citep{gayathri2020gw190521}. Due to the uncertainties in the mass distribution of astrophysical first generation BHs, we considered four mass distributions which obey the power law+peak model~\citep{2020arXiv201014533T}; but with different upper bounds. Our results demonstrate the importance of the upper limit of the mass distribution for the inference of the origins of BHs. 

For upper bound=$50M_{\odot}$ case, we find that GW190521 cannot involve two first generation BHs with certainty. Based on only time ordering and overlap of volume localization, for the non-eccentric localization we provide the two most plausible predecessors; GW190514 and GW170729. Both of the final masses of these mergers agree with the heavier mass of GW190521. In addition, the AGN from which a candidate optical counterpart was detected after GW190521 \citep{Graham_2020} is in the 90\% sky localization of GW190514. For the eccentric localization of GW190521, GW170823 has a better volume overlap than GW170729. However, the parameter estimation of the eccentric localization does not exist and based on the parameters of the best matching eccentric waveform (component masses=102 $M_{\odot}$) GW170823 does not seem as a plausible predecessor. These differences between the circular and eccentric estimations show the importance of the assumptions used in the detections for deducing the origin of the mergers. When we increase the upper bound to 60, 70, and 100 $M_{\odot}$, we find the GW190519-GW170818 pair to be most significant with $\sim 3\sigma$ significance. In addition to good overlap of volume and mass estimations of the merger pair, the mathcing BH of GW190519 is estimated to be spinning which is an expected characteristic of the triple hierarchical merger scenario and of hierarchical mergers in general. Despite the individual significances of the merger pairs, the whole search does not yield a statistically significant finding (except 50 $M_{\odot}$ upper bound) after the multiple hypothesis testing correction. Instead, the search provides interesting pairs which can be investigated in astrophysical contexts, i.e. GW190521-GW190514 pair as a hierarchical merger in an AGN \citep{2020arXiv201009765S,PhysRevLett.123.181101}, eccentricity of GW190514 or search for an AGN at the common localization of GW190519 and GW170818. Finally, in our test statistic we have not used the eccentricity as a parameter due to the lack of the proper eccentricity estimations for the mergers. With the development in the eccentricity analyses with produced waveforms, our search can be made more powerful by also using the eccentricity information, which is an expected characteristic in the triple hierarchical merger scenario \citep{10.1093/mnras/sty2249}.

% -------------------------------------------
\section*{Acknowledgements}
% -------------------------------------------
The authors thank Stephen Fairhurst for useful feedback. This document was reviewed by the LIGO Scientific Collaboration under the document number P2000459.

We acknowledge computing resources from Columbia University's Shared Research Computing Facility project, which is supported by NIH Research Facility Improvement Grant 1G20RR030893-01, and associated funds from the New York State Empire State Development, Division of Science Technology and Innovation (NYSTAR) Contract C090171, both awarded April 15, 2010. The authors thank the University of Florida and Columbia University in the City of New York for their generous support. The Columbia Experimental Gravity group is grateful for the generous support of the National Science Foundation under grant PHY-2012035. 

D.V. acknowledges support from Fulbright foreign student program and Jacob Shaham Fellowship. A.S. is grateful for the support of the Columbia College Science Research Fellows program. R.B. is supported by the School of Physics and Astronomy at the University of Birmingham and the Birmingham Institute for Gravitational Wave Astronomy. J.S. is supported by the
European Unions Horizon 2020 research and innovation programme under the Marie Sklodowska-Curie grant agreement No. 844629.

The samples for the IAS-Princeton detections were retrieved from \url{https://github.com/jroulet/O2_samples}.

We thank V. Gayathri et. al. for sharing the eccentric localization of GW190521.

This research has made use of data, software and/or web tools obtained from the Gravitational Wave Open Science Center (\url{https://www.gw-openscience.org/}), a service of LIGO Laboratory, the LIGO Scientific Collaboration and the Virgo Collaboration. LIGO Laboratory and Advanced LIGO are funded by the United States National Science Foundation (NSF) as well as the Science and Technology Facilities Council (STFC) of the United Kingdom, the Max-Planck-Society (MPS), and the State of Niedersachsen/Germany for support of the construction of Advanced LIGO and construction and operation of the GEO600 detector. Additional support for Advanced LIGO was provided by the Australian Research Council. Virgo is funded, through the European Gravitational Observatory (EGO), by the French Centre National de Recherche Scientifique (CNRS), the Italian Istituto Nazionale della Fisica Nucleare (INFN) and the Dutch Nikhef, with contributions by institutions from Belgium, Germany, Greece, Hungary, Ireland, Japan, Monaco, Poland, Portugal, Spain.

% -------------------------------------------
\bibliographystyle{aasjournal}
\bibliography{Refs}
\end{document}